*Review*

# Einstein, Barcelona, Symmetry & Cosmology: The Birth of an Equation for the Universe

Emilio Elizalde

Institute of Space Sciences (ICE/CSIC and IEEC), Campus UAB, Carrer de Can Magrans, s/n, Bellaterra, 08193 Barcelona, Spain; elizalde@ice.csic.es

**Abstract:** Albert Einstein visited Spain only once, precisely one hundred years ago. The circumstances, of a very different kind, of this visit will be explained here. In special, some important events happened to Einstein during that period, which, eventually, were key for converting modern cosmology into a genuine physical theory. Among them is the famous Einstein-Friedmann controversy, first, on the mathematical validity of Friedmann's equations and, later, their possible usefulness as a reliable tool to describe the real world. A summary of the deepest ideas underlying Einstein's contributions to the theory of relativity, which he had already completed before his visit, will precede the discussion, also supplemented with a description, in very simple terms, of the three main relativistic theories, namely Galileo's one, and Einstein's special and general theory. They pave the way towards a definitive theory of total relativity, so far unattainable. It will be recalled that the most general relativity principle, faithfully reflecting Ernst Mach's far-reaching ideas, might have much to do with the symmetry-breaking paradigm, a most crucial tool in quantum field theory and high energy physics.

## 1. Introduction

Albert Einstein visited Spain only once, which took place one century ago. This fact has been celebrated and extensively reported in the local and national media, in considerable detail, during the last couple of months. Surprisingly, however, some aspects of Einstein's visit went rather unnoticed, namely, the very important social and scientific circumstances surrounding his visit. When a journalist approached me last February, begging for brand new information about Einstein's stay in Barcelona and its surroundings, I had to think quite hard. What could I tell him that had not been written or said before in previous celebrations of the ephemerid? Nothing was my first reaction. But a minute later, I began to reconsider the situation. No doubt, 1923 was a glorious year for cosmology: the year of the famous Einstein-Friedmann controversy, which had started with the publication of Friedmann's fundamental equations in *Zeitschrifft für Physik* just a few months before the visit [1], and which ended a couple of months after it, with Einstein's acceptance of Friedmann's formulas, which give a faithful description of the universe we live in. Friedmann's equations are now recognized as the formulation of Einstein's general theory of relativity [2] that correspond to our universe and constitute the basis of all modern cosmology. More specifically, we have one general equation (Einstein's) with just one valid solution (Friedmann's) for ruling the whole cosmos: a unique, incredible achievement in the history of cosmology and, by extension, in all of Human History.

This issue was paramount in establishing cosmology as a modern science [3,4]. And all this happened around 1923, quite precisely, around the time of Einstein's visit to Spain. At a second instance, having to prepare the introductory talk for the 4th Symmetry Conference, I was brought to dig more deeply into Einstein's relativity theory—which was the main subject of all of Einstein's talks at that time—and then suddenly recalled the following important fact: the relativity principle may indeed have much to do with the symmetry breaking paradigm! Thus, I found a strong connection among all variables and was ready to write a report offering a novel approach to the subject.

In what follows, an exposition will be made, albeit necessarily limited, of the scientific context and, more generally, of the historical, economic, and social circumstances corresponding to the epoch of Einstein's trip a century ago. In particular, the environment and the general circumstances of a very wide scope in which the visit took place will be considered. Not many new or more precise details of his stay will be given; only a few corrections to previously issued, inaccurate statements on the ephemerid.

The contents of the paper are as follows. In Section 2, a summary of the main results obtained by Albert Einstein before 1923 will be discussed. Special emphasis will be made, in Section 3, on the essential principles conforming to his two famous theories of relativity, namely the special and the general one; both will be described as extensions of the pioneering relativity (or covariance) principle due to Galileo Galilei, and as (first- and second-order) attempts to crystallize the very ambitious ideas formulated by Ernst Mach. Section 4 will recall the main events that occurred in the world in 1923 and, subsequently, Einstein's six-month-long journey, which started on 6 October 1922. The essential scientific context surrounding Einstein during his trip, which ended with his visit to Spain will form the content of Section 5, including the famous Einstein-Friedmann controversy. The paper ends with Section 6, containing some conclusions and an outlook.

## 2. Who Was Einstein? What Had He Achieved by 1923?

As is well known, Einstein's most prolific year in his entire life was 1905, when he was just twenty-five and working at the patent office of the Swiss Federal Institute for the Intellectual Property in Bern. That year is sometimes described as his *annus mirabilis* ('year of miracles', no wonder that 2005 was declared the World Year of Physics), during which Einstein published four extraordinarily momentous and groundbreaking papers [5]. Many scholars claim that each one of these papers could have deserved the Nobel Prize. In one of them, he established the theory of the photoelectric effect; in another, he explained the elusive concept of Brownian motion; and, in the last two papers, he introduced the theory of special relativity and demonstrated the daring mass-energy equivalence, which was to have so many implications for the future of humankind (although he did not explicitly write down, in that paper, his most famous formula, yet).

Einstein observed that the laws of classical mechanics did not agree with those of the electromagnetic field, which led him to develop his particular theory of relativity. It took him, however, another ten years, and many more efforts, to extend that theory to the gravitational field; until he arrived at his theory of gravitation, or general theory of relativity. The consequences of the latter went far beyond those of Newton's laws, including his universal law of gravitation.

As verifiable evidence that Einstein's general relativity (and not Newtonian mechanics) was the correct theory—and as proofs that could serve to establish clear observational evidence of its validity—Einstein referred to the anomalous precession of Mercury's perihelion, to the deflection of light in gravitational fields, and the gravitational redshift. His general relativity made precise numerical predictions about these three effects that differed from the results obtained in Newtonian gravitation.

Already in 1915, Einstein could calculate with his $\sqrt{\ }$ equations—although approximately using them— the anomalous precession of Mercury's perihelion (an effect much easier to obtain from Schwarzschild's solution, which he still did not have [3]). And he found a value that perfectly matched the observations of the anomaly, which was a significant issue at the time. Deeply excited, he hurried to communicate the good news to his friend Michele Besso. Einstein had already become fully confident that his theory was correct!

And this happened four years before the famous observation of the solar eclipse of 1919, which constituted for the rest of the world the definitive confirmation of Einstein's theory, beating Newton's. This fact, almost incredible at that time for everyone, was published on the front pages of all newspapers and magazines everywhere and made Einstein a world-famous person. Until then, nobody had even imagined that Newton would ever be challenged. For Einstein himself, on the contrary, there was no surprise. When he was asked by some journalist in 1919 what his reaction would have been if the data obtained from the solar eclipse had not confirmed his calculations, he answered, without hesitation, that if that had been the case, then "the error would necessarily have been in the observations of the eclipse, since I have for certain that my theory is correct". It must be noted that, in both cases, the relativistic contribution to the corresponding effect is not minor. On the contrary, it is of the same order of magnitude as the classical effect. This goes against the often belief that general relativistic effects are small with respect to Newtonian ones.

Going back in time, it was in 1917 when Einstein first applied his general theory of relativity to the description of the Universe. And, as he immediately saw that he had the same difficulty as Newtonian physics in modeling a static universe, he had no choice but to introduce a universal constant: the cosmological constant. In a similar way, as he explains in all detail in his work [6], as it can be done in the Newtonian case. And Robert Hooke, who has sometimes been termed "the genius in the shadow of Isaac Newton", had already considered as a possibility several centuries before [7]. Indeed, in that respect, the problems of Newtonian gravity and general relativity are the same and the 'solution' to render a static universe (by means of the cosmological constant), too. But this solution turns out to be unstable and, therefore, not useful, as was later discovered (by Eddington and Lemaître, among others [3]).

The ones mentioned above had been Einstein's most important discoveries before he visited Spain and had already earned him several nominations for the Nobel Prize in physics. He had long been convinced he would be granted the prize [8]. When he was finally awarded, in a thank-you letter he wrote to the Nobel Committee, he joked that "he was very happy to have received the award, in particular, because, from then on, he would get rid of so many boring people who kept asking him all the time, how it was that he had not yet got the award". He finally got it in 1922, although the prize did correspond to 1921, a year in which the objections of some members of the Committee had prevented it from being granted to him for his theory of relativity. As we will later see, when he visited Spain, he had not yet had the opportunity to make the prescriptive speech at the formal acceptance ceremony, which did not take place until July 1923.

The Nobel Committee's task had not been simple, indeed. The creation of the Nobel Prize was still quite recent; there were few precedents, and Alfred Nobel's will established that the prize was to be awarded to *"those who, during the previous year, have conferred the greatest benefit on humanity"*. At that time, it was unclear how important Einstein's work was for humanity. And, as for the general theory of relativity, only a few true specialists understood what he had done. It is therefore not surprising that, finally, the Committee decided to award him the prize (on reconsidering the case in 1922) *"for his services to theoretical physics, and especially for his discovery of the law of the photoelectric effect"*. His law of the photoelectric effect had already been verified experimentally in 1916 by Robert Millikan, who would receive the prize the following year, 1923.

Einstein was very confident that he would eventually be awarded. The surprising point is that in the diary he kept, quite schematic but very detailed, Einstein did not mention the day he knew he had won the prize! And another equally remarkable fact is that, in the separation agreement of his first wife, Mileva Maric, which took place in 1918, Einstein offered her a monthly income associated with the award, for her and their two children (one of whom required very expensive medical attention), in the case, he would get it. Mileva accepted such an agreement, showing that both considered this a highly plausible possibility. The material value of the prize was equivalent to about fifty times Einstein's annual salary, which, in a short time, was significantly devalued (along with the monetary deposits he might have had) due to the economic situation in Germany after the Great

War. We will deal with this and other crucial circumstances of the epoch in Section 4 while approaching the time of Einstein's visit chronologically. But first, some important concepts.

## 3. Essential Principles Conforming the Theories of Relativity

On 25 November 1915, in his intervention at the Prussian Academy of Sciences session, entitled *"Die Feldgleichungen der Gravitation"*, Albert Einstein announced his General Theory of Relativity, on which he had been working tirelessly for nearly ten years. A year and a half later, on 8 February 1917, in another speech at the same Academy—this one with the title *"Kosmologische Betrachtungen zur allgemeinen Relativitatstheorie"*—he applied his new theory, for the first time, to the description of the universe. In this Section, the critical significance of these facts will be discussed, among other early episodes—also very relevant and corresponding to astronomical observations—which culminated in the birth of Modern Cosmology. This discipline eventually took, as a solid theoretical basis, the so-called field equations of the General Theory of Relativity.

The present discussion's originality relies on the fact that it has for a reference and guiding thread the very important scientific, social, and economic environment of the above-mentioned Einstein's visit to Catalonia and Spain, of which we are now celebrating the centenary. It will be noted, in particular, that one of the most important episodes of the conversion of Cosmology into a modern science occurred precisely around that trip, a crucial fact that tends to go completely unnoticed everywhere.

One of the questions I was asked at a round table celebrated on the occasion: *"Giving curiosity a voice"*, an event commemorating the mentioned ephemerid and organized by the Catalan Foundation for Research and Innovation (FCRi), with the collaboration of Divulcat and Astro Barcelona, was: please continue this sentence *"The theory of relativity is based on..."*. A short answer had to be provided in just a few minutes, so I did it on the spot. Here, we will have more lines to complete in more detail.

*3.1. Galilean Relativity*

The remarkable Galileo Galilei (1564–1642), considered the founder of modern science (Figure 1), was the first to formulate a principle of relativity, or covariance, in an evident and beautiful way. This principle clearly expresses the very important fact that "it makes sense to talk about laws of physics"; that is to say, these laws do not change. They are immutable when we move from here to any other place in the universe or get on board a vehicle that moves in a straight line and at a constant speed. This is called an inertial frame that will remain the same forever if no force is acting.

Galileo, in his famous book of 1632 *"Dialogo sopra i due massimi sistemi del mondo"* masterfully expressed this principle in the words of Salviati, when he proposes (on the second of the four days of dialogues) the following experiment [silence, please, it is Galileo himself who speaks to us] [9]:

> *"Lock yourself up with a friend in the main cabin, under the deck of a rather large ship, and bring flies, butterflies, and other small flying animals. Hang a bottle so that it drains, drop by drop, into a large container below. Make the ship go at the speed you prefer, but always the same: a smooth motion without fluctuations in one direction or the other. The drops will fall into this container without being diverted aft, even if the ship has moved forward while the drops are still in the air. The butterflies and flies will continue their usual flight from side to side as if they never tire of following the ship's course, however fast it may go, and it will never happen that they concentrate on the stern of it."*

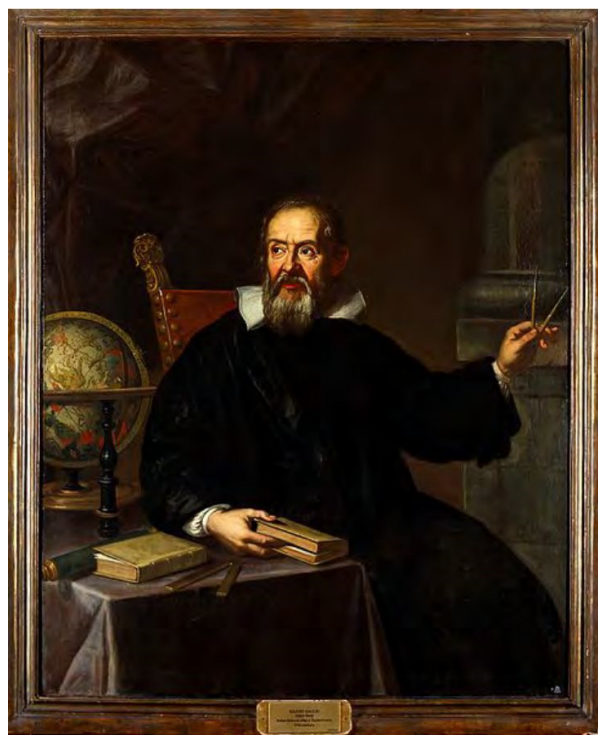

**Figure 1.** Galileo Galilei (1564–1642). Oil portrait of an Italian painter believed to be from the 18th Century. Reproduced with permission from Wellcome Trust, UK. Fair use.

It is certainly an accurate and most precious description of the principle of relativity. To get it right, although the laws of physics do not change when passing to a different system, their specific manifestation changes. As the law itself, the mathematical equations expressing the law remain the same, but the solutions look different in different frames and are connected by a Galilean transformation. In other words, the evolution of a particle, i.e., its world-line, is different since the initial conditions are different in different inertial reference frames. This is taken care of by applying a transformation of what is now known as the Galilean group, in order to go, in the case considered, from the reference system located on land to the reference system fixed on the ship, or vice versa. In the latter frame, the ship remains still, and it is the sea around it and the port from which it set sail that is constantly moving. This is where the name relativity comes from: the description of each reference system is different, although the physical law, the essence, is unaltered.

Galilean relativity is the simplest of all relativistic theories. Galileo had great intuitions that he linked with observations of nature and some experiments that he carried out personally (although there are still discrepancies about how many of them, he did, in reality). He lacked knowledge of mathematics, which he proclaimed was the language in which the laws of nature should be written. But Galileo used geometry almost exclusively. In the opinion of the great Nobel laureate Steven Weinberg, if he had used more algebraic tools, he could have gone much further [10]. The fact that he could formulate so accurately, in plain words, the principle of relativity proves that this statement is not misguided.

Fifty years ago, Jean-Marc Lévy-Leblond devoted his efforts to rescuing and polishing Galilean relativity, formulating it mathematically like Einstein's special relativity. An elaborate and beautiful theory emerged from his work [11,12], parallel to Einstein's much more famous theory, to be discussed below. There is only one difference, apparently small but essential: the constancy of the speed of light, $c$, in any inertial frame of reference, implying that its value can never be exceeded (for transmitting information of any kind).

We will end this summary of Galileo's relativity with two comments. The first Weinberg's criticism of Galileo could also be extended to Isaac Newton (1642–1727) who, despite being the creator—together with Wilhelm Leibniz (1646–1716)—of the very powerful in-finitesimal calculus [13–15], never used it, practically, in the formulation of the laws of his mechanics, which in the Principia are given in the form of endless paragraphs challenging to digest and to use.

The second comment concerns my work on this topic, particularly the one I carried out for my PhD thesis [16–18]. Starting from the papers of Lévy-Leblond and other authors, we came to connect, both ways, the theory of Lie groups of the respective transformations: using techniques of contraction and dilation of groups in changing dimensions, we related the groups of Galileo and those of Lorentz and Poincaré, which correspond to the special theory of relativity [19–22] (Figure 2). It is not time to go deeper into these concepts. Still, it must be emphasized that all these developments have given even more relevance to the ideas and formulations of Galileo as the true pioneer of relativistic theories.

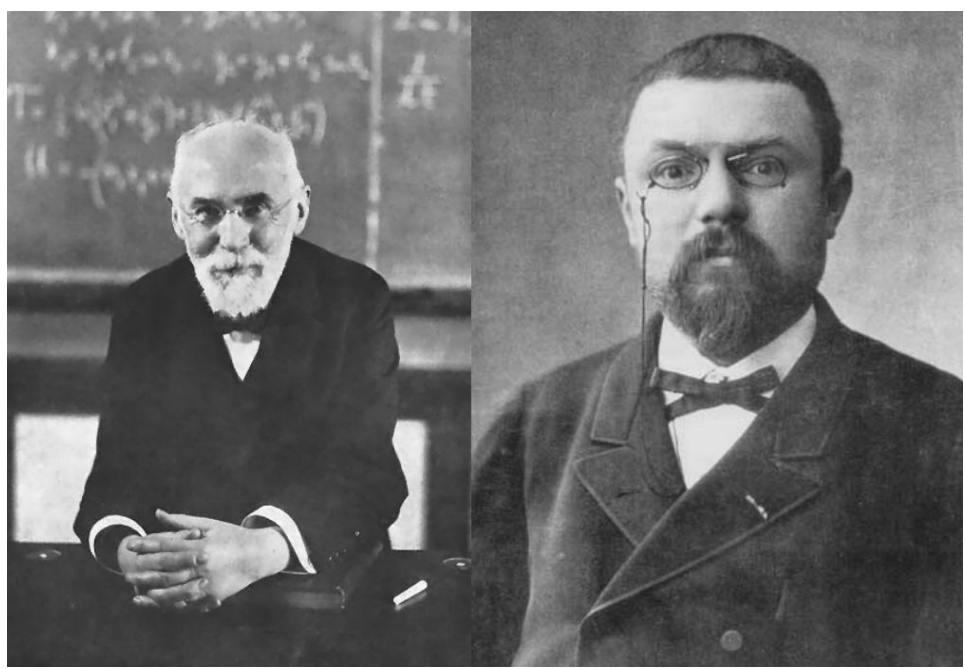

**Figure 2.** On the left, Hendrik Lorentz (1853–1928) and, on the right, Henri Poincaré (1854–1912)—Image: Wikimedia Commons. Public Domain.

*3.2. The Special Theory of Relativity*

Towards the end of 1905, in one of the four historical works written during what is often called his *annus mirabilis*, Albert Einstein published his special theory of relativity (Figure 3). Masterfully, in one of these works, he was able to derive the Lorentz transformations under only two assumptions of the principle of relativity or covariance (Galileo's one, which we have already seen) and of the constancy of the speed of light (in ideal vacuum conditions) in any inertial reference system—a fact that had been already checked in the famous experiment of Michelson and Morley—and at the same time abandoning the ether as just unnecessary [23].

In this way, Einstein filled with meaning the Lorentz transformations (consisting of rotations and displacements at constant speed) and the Poincaré ones (also including space translations), which had been previously considered by different physicists since 1887. All these transformations reduce to those of Galileo when the speed between the two reference systems is much smaller than that of light.

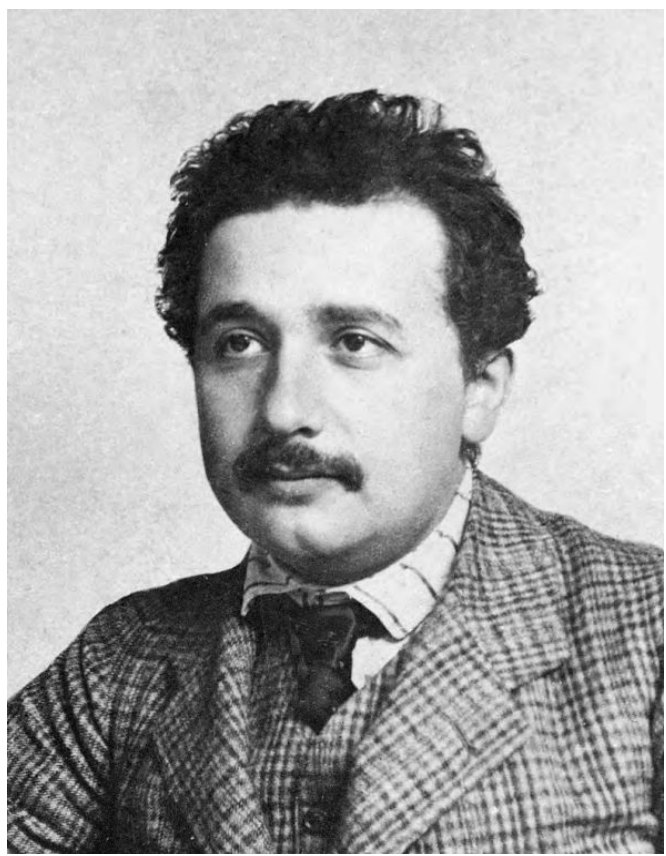

**Figure 3.** Albert Einstein, around 1905, his "annus mirabilis", in which he published four momentous articles. One of them: "Zur Elektrodynamik bewegter Körper", the work in which he built the special theory of relativity. Public Domain.

Summing up, regarding the postulates, in his special theory of relativity, Einstein only added to the principle of relativity due to Galileo a second postulate, which states that the speed of light in a vacuum is the same for any inertial reference system. The consequences of these two simple postulates are amazing and very difficult to understand by those of us who always move at insignificant speeds compared to light. Completely improbable phenomena appear, even seemingly absurd situations, such as the fact that the simultaneity of two events is relative (to the reference system), the phenomena of time dilation, length contraction, a relativistic contribution to the Doppler effect, and many other weird phenomena.

They are immediate consequences of the two postulates and are obtained simply using the corresponding Lorentz transformation. It is true that they only manifest themselves when the speed at which one system travels concerning the other is close to that of light, but it must be observed that this condition already occurs nowadays in a multitude of laboratory experiments carried out with elementary or just very small particles, also in photonics, and at very different levels (think of the ubiquitous GPS signals, which we are using all the time [24]).

But maybe the most extraordinary consequence for human society that Einstein's special theory of relativity had was the realization of the equivalence between mass and energy, very simply expressed by his most famous formula: $E = mc^2$. Einstein took time to write it in this form; he did not do so in his already-mentioned work of 1905, in which he expressed it indirectly, already. In principle, the formula describes the values that the magnitudes take in a reference system at rest, but it also extends to the values of relativistic mass and energy for a system in motion. Einstein clearly stated that the laws of energy conservation and conservation of mass were *"the same"* [25].

Anyway, some physicists consider this formula to be overrated, as it has little real utility in designing how to carry out, in practice, nuclear fission processes. Be that as it may, it is very accurate that Einstein's formula served as a guide when evaluating this possibility for the first time in history, namely, to understand whether nuclear fission had occurred in a certain laboratory experiment. We have a magnificent, first-hand description of how this occurred, in the words of Otto Frisch, one of the main actors of this play.

During the Christmas holidays of 1938, Frisch spent some days in Stockholm at the invitation of his aunt, the great Lise Meitner. One day, both went out for a walk in the cold of the snowy city. While talking at length about the issue, they managed to understand the meaning of the experimental results of their colleagues Otto Hahn and Fritz Strassmann in Berlin. By bombarding uranium atoms with neutrons, those had obtained what appeared to be barium and an excess of neutrons, a completely unexpected and mysterious result. Frisch and Meitner understood that the uranium nucleus had been split and introduced the idea of what would later be called atomic fission. They directly used Einstein's equation to quantify the energy of a reaction that should be able to overcome forces such as the surface tension, which holds the nucleus together, to allow the fission fragments to drift apart a little, resulting in a configuration from which their charges could force them (by electrostatic repulsion) into an energetically favorable final fission. Using the packing fraction, or value of nuclear binding energy per nucleon, along with the formula $E = mc^2$, they realized that the basic process of fission *"was indeed energetically possible"*. As Frisch described it [26]:

> "We walked up and down the snow, me on skis and Lise on foot...and little by little the idea took shape... based on Bohr's conception of the nucleus as a drop of liquid; the drop could stretch out and split apart... We knew there were very strong forces that would oppose it,... like the surface tension. But nuclei are different from normal droplets. At this point, we both sat on a tree trunk and started calculating on scraps of paper... the uranium nucleus could become a very unstable blob, ready to split... But... when the two blobs separated, they would be further separated by electrical repulsion, the equivalent (in energy) of about 200 MeV. Fortunately, Lise remembered by heart how the masses of nuclei were calculated... and discovered that the two nuclei formed... would be lighter by about one-fifth the mass of a proton. Now, every time mass disappears, energy is created, according to Einstein's formula $E = mc^2$, and... the loss of mass was equivalent to 200 MeV! Everything fit!"

Fission could therefore take place! When, sometime later, Einstein found out that it had been carried out and its dramatic consequences, it is said that he exclaimed: *Woe is me!*

*3.3. The General Theory of Relativity*

Regarding the special, it contains only one additional postulate, the principle of equivalence. Einstein formulated it one day after having what he would later describe as *"the happiest idea of my entire life"* (other authors place this famous sentence at the moment he found the formula we discussed above).

It was Einstein who explained (although there is no written record of it) that the idea came to him in 1907 while working at the Patent Office in Bern. He was sitting in his usual chair in front of his desk when suddenly he was very startled by a thought that occurred to him about what would happen if, at that very moment, he fell upright from the roof of his house. He continued to reason slowly... At that instant, while he was falling, no gravitational field would exist for him as an observer, at least not in his surroundings. Indeed, if he had an object in his hand, say an apple or a coin, and simply let it go, the object would not fall at his feet; it would always remain next to his hand without separating from it: it would not experience, therefore, any gravity! If he could not see anything except himself and the object, he might reasonably conclude that he was in a zero-gravity place. Later, the example of an elevator in free fall with a person inside has been commonly used as an alternative to illustrate the same idea (in this case, the elevator walls already isolate the experimenter from the rest of the world).

Expressed in another way, the conclusion is that the force of gravity is not special at all: it is just like any other mechanical force that sets an object in accelerated motion. Another alternative version of the same principle is to consider that the mass of a body involved in Newton's formula of universal gravitational attraction (the gravitational mass, $m_g$) is the same as the one which appears in the formula $F = m_i a$ (called inertial mass, $m_i$), which is inversely proportional to the acceleration that the body acquires when a mechanical force is applied to it. In short, $m_g = m_i$. All these formulations of the principle of equivalence are equally valid.

That thought of Einstein (one of his most famous *gedanken* experiments) was quite happy since it led him to build a whole new theory of gravitation, which he called the general theory of relativity and has gone much further than Newton's universal gravitation. In it, as was already the case for special relativity, space and time are united in a continuous "fabric" of space-time; but, as a great novelty, the presence of matter now results in a local curvature of this fabric, similar to what happens when a child gets ready to jump on an elastic bed, at a fair, thus making the fabric, originally flat, to collapse under its weight. In this theory, the curvature of space-time gives rise to the effect we call gravity.

In more technical terms, Einstein's equivalence principle for a uniform gravitational field states that the motion of an object in an inertial frame of reference is indistinguishable from the motion of the object in the absence of that field but concerning a suitably uniformly accelerated reference system. In his own words (Einstein, 1907) [27]:

> "*We assume the complete physical equivalence of a gravitational field and a corresponding acceleration of the reference system*".

Continuing with the "happiest thought of his life", Einstein also referred to two reference systems, K and K'. K has a uniform gravitational field, while K' has no gravitational field but is uniformly accelerated so that the objects in the two systems experience identical forces (Figure 4). Again, in his own words (Einstein, 1911) [28]:

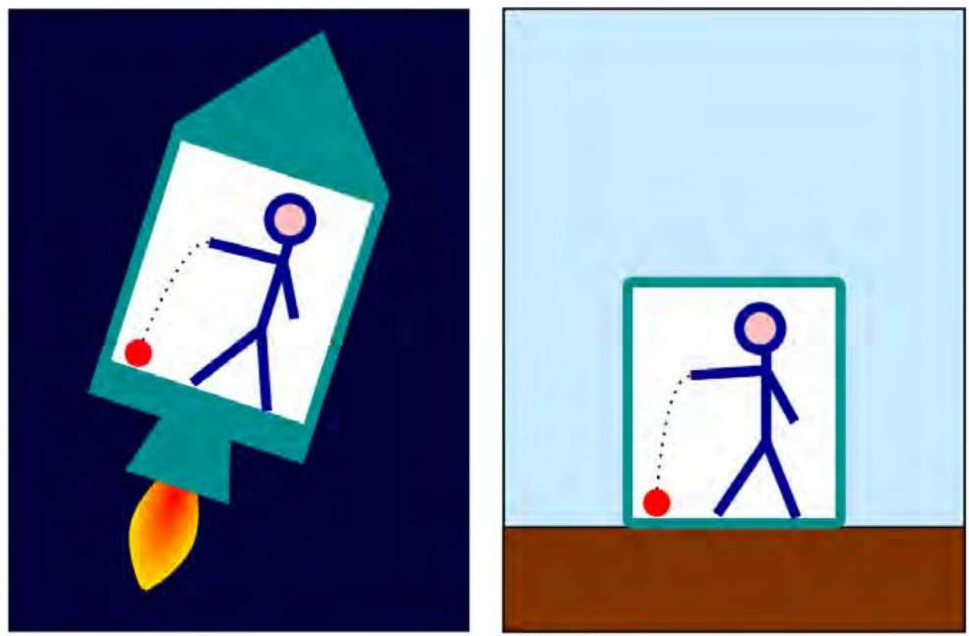

**Figure 4.** On the (**left**), a ball falls to the ground in a suitably accelerated rocket in the absence of gravity. On the (**right**), the ball falls to the ground in the usual way. The effect is identical in both situations, completely indistinguishable in the chamber that isolates the observer from the outside world. Fair use.

> *"We arrive at a very satisfactory interpretation of this law of experience if we assume that the systems K and K' are, physically, completely equivalent; that is, if we admit that we can also consider the system K as a space free of gravitational fields, but at the same time as a uniformly accelerated system. This assumption of exact physical equivalence makes it impossible for us to talk about the absolute acceleration of the reference system, just as the theory of special relativity forbids us to talk about the absolute speed of a system. And this makes the equal fall of all bodies in a gravitational field seem then to be a most natural thing."*

However, we should not be fooled by the apparent simplicity of all these concepts. Here we are talking only about the fundamental principles of the general theory of relativity, but it is necessary to mention that it took Einstein ten full years of his life, working without rest, to arrive at his final field equations from these principles. This would give to another article, quite long and much more complex. Here, we will limit ourselves to writing his field equations.

$$R_{\mu\nu} - \frac{1}{2}Rg_{\mu\nu} + \Lambda g_{\mu\nu} = 8\pi T_{\mu\nu},$$

where $R_{\mu\nu}$ is the Ricci curvature tensor, which represents the curvature of spacetime caused by matter; $R$ is the scalar curvature, a measure of the overall curvature of spacetime; $g_{\mu\nu}$ is the metric tensor, which describes the geometry of spacetime; $\Lambda$ is the cosmological constant, which in its modern conception represents the energy density of the vacuum state of space; and $T_{\mu\nu}$ is the stress-energy tensor, which describes the distribution of matter and energy in spacetime. The Ricci curvature tensor is a geometric object obtained by contraction of the first and third indices of the Riemann curvature tensor, $R^\rho{}_{\sigma\mu\nu} = \partial_\mu \Gamma^\rho{}_{\nu\sigma} - \partial_\nu \Gamma^\rho{}_{\mu\sigma} + \Gamma^\rho{}_{\mu\lambda}\Gamma^\lambda{}_{\nu\sigma} - \Gamma^\rho{}_{\nu\lambda}\Gamma^\lambda{}_{\mu\sigma}$, namely $R_{\mu\nu} = R^\rho{}_{\mu\rho\nu}$, where $\partial_\mu$ stands for $\partial/\partial x^\mu$, $\Gamma$ are the Christoffel symbols of the Levi-Civita connection corresponding to the spacetime metric, to wit: $\Gamma^\rho{}_{\mu\nu} = 1/2\, g^{\rho\sigma}\,(\partial_\mu g^{\sigma\nu} + \partial_\nu g^{\sigma\mu} - \partial_\sigma g_{\mu\nu})$. One must observe that Einstein only introduced the cosmological constant term in his paper of 1917, *"Kosmologische Betrachtungen . . ."* [6], where he used his field equations for the first time to obtain a static model for the universe.

We cannot refrain from bringing up a passage where Einstein commented on some aspect of math necessary to translate his principles into useful formulas and equations. In a letter to Arnold Sommerfeld from the year 1912 (that is, about sixty years after Bernhard Riemann's famous habilitation work) [29], Einstein commented on the efforts he was making to learn Riemannian geometry. He says:

> *"Aber eines ist sicher, dass ich mich im Leben noch nicht annähend so geplagt habe und dass ich große Hochachtung vor der Mathematik eingeflößt bekommen habe, die ich bis jetzt in ihren subtileren Teilen in meiner Einfalt für puren Luxus gehalten habe!"*

Which translates to:

> *"But one thing is certain, that I have never nearly so toiled in life and have instilled in me such a high regard for mathematics, which I had considered until recently, in my naivety, as for its subtler parts, as a simple luxury!"*

Going a little deeper into the principle of equivalence, three forms are currently being considered: the weak (or Galilean), the Einsteinian and the strong equivalences. In the weak equivalence principle, also known as the universality of free fall or Galilean equivalence principle, the universality of free fall is restricted to ordinary bodies (the ones Galileo had in mind only), which are bound just by non-gravitational forces (for example, a stone, a piece of metal, a block of wood, etc.). We thus see that Galileo was also a pioneer in the conception of an equivalence principle, which Einstein would extend in his general theory of relativity.

In his precise words, the Einsteinian form of it is the one we have already considered above (Figure 5). And the principle of strong equivalence is a generalization of the two, which also includes as bodies the astronomical objects, such as pulsars and black holes, which are very unusual, for they have a very important part of the energy that holds them together being of gravitational type. Strong equivalence can be tested by looking for a variation in Newton's gravitational constant, G, or a variation in the masses of the fundamental particles over the universe's lifetime. Observations of an independent nature, such as precision measurements of Solar System orbits and studies of Big Bang nucleosynthesis, have shown that G cannot have varied by more than 10% along time. The strong equivalence principle can also be checked by looking for some kind of fifth force, in terms of deviations from the gravitational law predicted by general relativity, as searching for inverse square law errors, in terms of Yukawa forces or violations of Birkhoff's theorem.

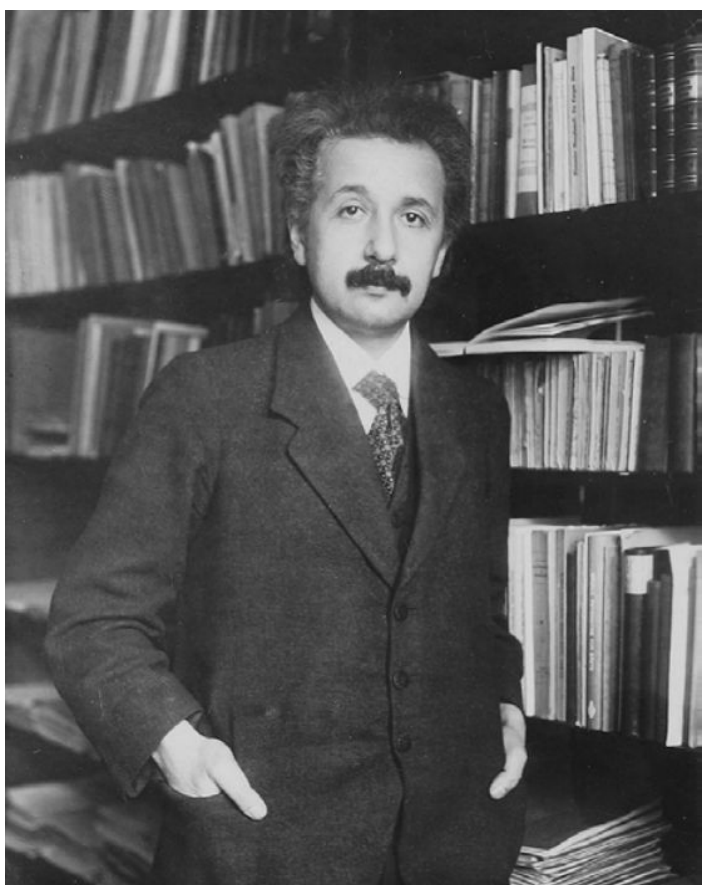

**Figure 5.** Albert Einstein in 1916, in the house library by Paul Ehrenfest, in Leiden, where he stayed for a few days. Public Domain.

And with that, we have reached the more than reasonable ceiling of what may be explained in a short introduction, such as this one, to the theories of relativity.

Just for completeness, as has been mentioned already, the first exact solution to Einstein's field equations was found by Karl Schwarzschild in 1916, only a few weeks after Einstein presented them to the Prussian Academy. The Schwarzschild solution reads:

$$ds^2 = (1 - 2GM/rc^2)\, c^2 dt^2 - dr^2/(1 - 2GM/rc^2) - r^2(d\theta^2 + \sin^2\theta\, d\varphi^2).$$

It is an exact solution that describes the gravitational field outside a spherical mass on the assumption that the electric charge of the mass, its angular momentum, and the cosmological constant are all zero. It is a useful approximation for describing slowly rotating astronomical objects such as usual stars and planets, including our Earth and the Sun.

For future reference in the present article, let us also write down the Friedmann–Lemaître–Robertson–Walker (FLRW) metric, which describes a homogeneous and isotropic universe. In reduced-circumference polar coordinates, the metric has the form:

$$ds^2 = -c^2 dt^2 + a(t)^2 \frac{dr^2}{1 - kr^2} + r^2(d\theta^2 + \sin^2\theta d\phi^2).$$

Here, a(t) is known as the scale factor, k is a constant representing the curvature of the space, and r is sometimes called the reduced circumference, which is equal to the measured circumference of a circle (at that value of r), centered at the origin, divided by 2π (like the r of Schwarzschild coordinates)

Before ending this Section with a telegraphic summary of everything exposed above, the following, most relevant observations are in order.

*3.4. Two Important Observations*

First. The first and third principles, general relativity or covariance and equivalence, are the two basic postulates of the theory of general relativity (apart from the second, equally key but already inherited from special relativity). There has been a lot of discussion about its independence or not. The answer is not so immediate. The two principles are usually presented as independent, and they are quite so in their most strict formulation. But it happens that, in practice, they are connected by inaccuracies that come from the approximations made in formulating the equations of general relativity, which, for simplicity, Einstein reduced to the second order. This means that the equivalence principle is approximate (in its implementation in Einstein's equations) and valid only for second-order terms (accelerations). Indeed, the accelerations are indistinguishable at any given point; but the differentials of these and the gradients of higher orders are not identical. And this error in the equivalence (although very small in practical situations) introduces problems with the covariance, which causes the terms higher than the second order to be truncated. In short, the curvature of space is well represented in the equations, but not higher-order spacetime deformations.

Einstein was the first to admit that his final theory was approximate and incomplete. He hoped other scientists would improve it soon, which has not happened, not even now, although many have tried, and many alternative theories have certainly appeared in the meantime. General relativity works very well up to high values for the energy: it has recently done so to high accuracy in describing black hole collisions of about thirty solar masses and more. But if the kinetic energy were even much larger, high enough to fold space-time into layers, then the problems would already appear in a clear way. Among several others, a candidate aiming to improve the theory at much higher energies is topological geometrodynamic (TGD), a theory involving highly complex mathematics [30].

Second. Einstein's attempt to somehow materialize the ideas of Ernst Mach in his construction of general relativity was undoubtedly a very important stimulus for creating his theory. However, it does not appear as one of the fundamental principles of the same (Figure 6). The name Einstein gave to his theory came from his conviction that he could do justice to Mach's criticism of Newton's notion of absolute space. According to Mach, space had to be relativistic (or covariant) concerning the most general possible transformations of the spacetime coordinates.

Mach's principle of total relativity goes beyond the equivalence principle and, according to the Nobel Prize winner Frank Wilczek (from whom I have taken the liberty of borrowing some of these sentences), it may be understood at the same time as a principle of symmetry. In the primary equations, we should put all possible movements on an equal footing, not just those related by a constant speed or constant acceleration (which remain as first- and second-order approximations to such a goal). Any choice of coordinates should be equally valid since the coordinates should be subject to arbitrary variations. General relativity always includes a metric field, which tells us how to assign numerical measurements to intervals of time and space, and one always tries to choose schemes in which the metric field takes the simplest possible form. But we should be able to place ourselves further, before choosing this field, before the actual specification of space and time. This is demanded, in special, if we want to approach the very origin of the universe. But, again, this would take us too far.

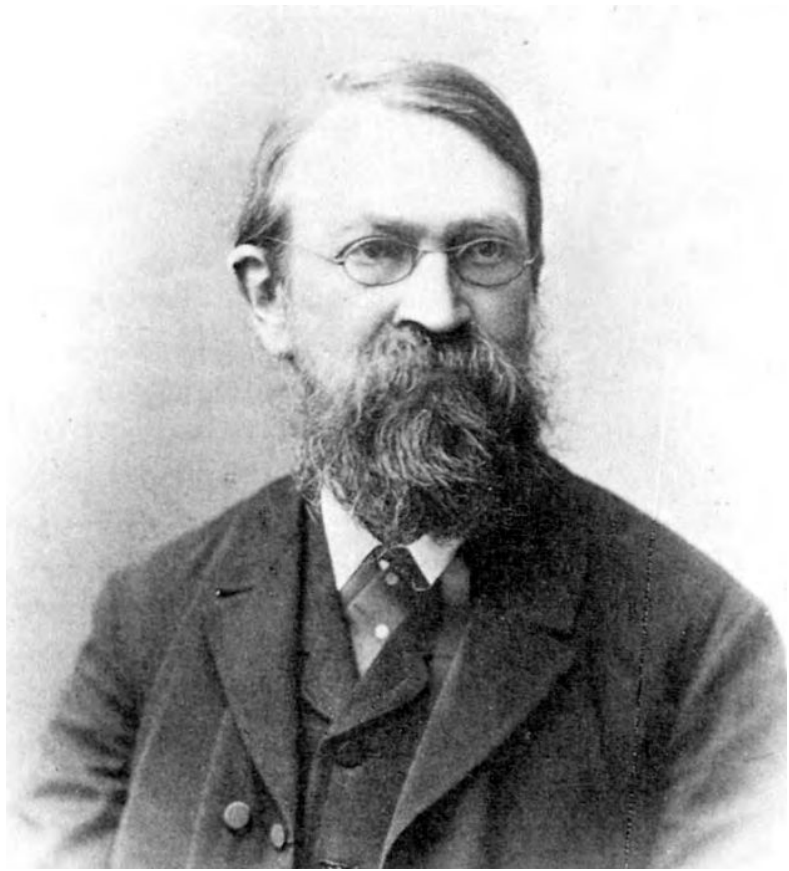

**Figure 6.** Ernst Mach, Austrian physicist, and philosopher (1838–1916)—Public domain.

To conclude, when the problem is cast in this way, it is very similar to the concept of symmetry breaking, so crucial in the modern formulations of theoretical physics. Recall, for example, that in the electroweak standard model, a Higgs field appears that breaks the symmetry of the primary equations and gives mass to the elementary particles; on the other hand, in quantum chromodynamics, a quark-antiquark condensate arises that also breaks the symmetry; and in the schemes of great unification, several generalizations of the same idea are used. The symmetry perspective would contemplate the possibility of primary theories enjoying greater symmetries than those realized in the equivalence principle of general relativity. In this context, Mach's principle would be the hypothesis that the most general primary theory should include the principle of total relativity, that is, the physical equivalence among all possible coordinate systems (Wilczek, 2004) [31]. Again, we have flown too far: no one has yet advanced substantially, in practice, along this path.

*3.5. Summary of the Relativity Theories*

In summary, the three fundamental principles of the General Theory of Relativity are as follows (Figure 7):

1. The principle of relativity or general covariance
    - Galileo: it makes sense to formulate laws of physics (inertial systems)
    - Laws for inertial or accelerated systems. Form of eqs changes (Galilean, Lorentz, Poincaré transf.)

- No total relativity (Mach's principle). Truncated to 1st-/2nd-order eqs.
2. The speed of light in a vacuum is constant, c, in all inertial systems
    - Together with Galileo's relative principle (inertial system) → Special Theory of Relativity
3. The principle of equivalence
    - Gravity is like all other forces. Equiv. of inertial mass and gravitational mass: $m_i = m_g$
    - Spacetime is a mathematical manifold, locally Minkowskian

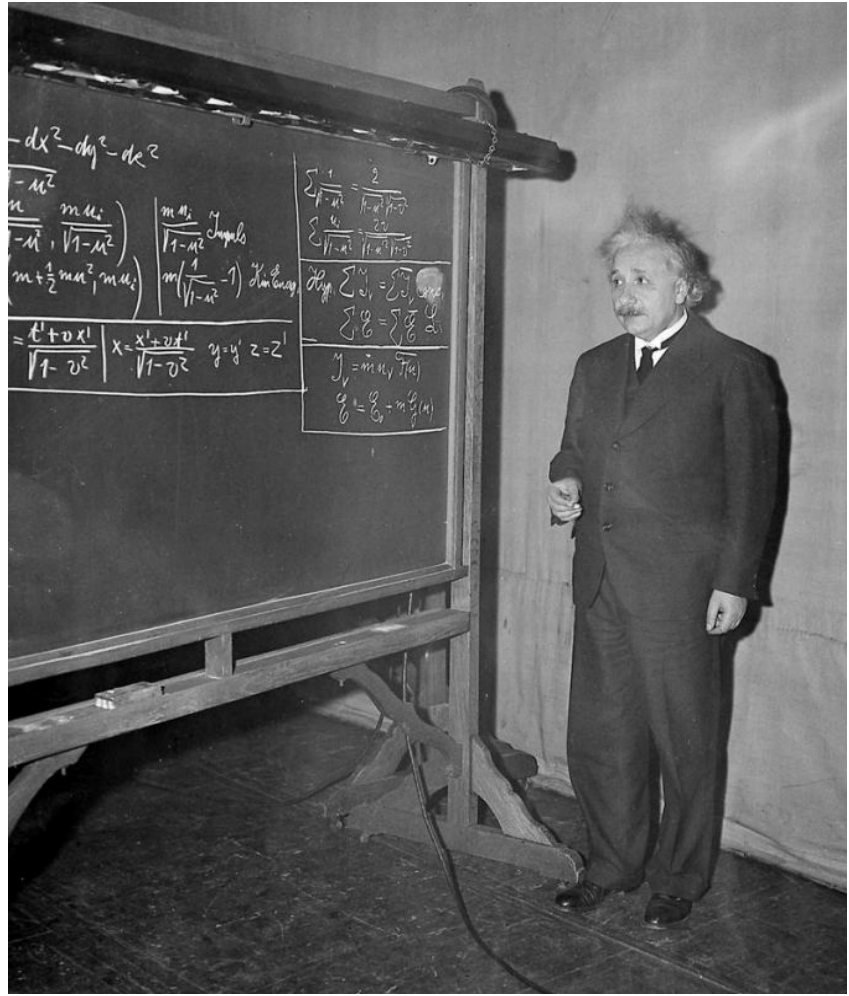

**Figure 7.** Albert Einstein giving the 11th Josiah Willard Gibbs lecture at the American Association for the Advancement of Science meeting on 28 December 1934—Public domain.

They are completed with two more conditions of a technical nature and refer to the field equations of the theory.

4. Zero torsion hypothesis ($\nabla_X Y - \nabla_Y X = [X,Y]$)
    - Christoffel symbols are symmetrical. You can relax it (Einstein-Cartan, string theory)
5. Reduction to Newton's laws (at small speeds as compared to c)
    - To define the universal constants of the new theory

And a summary of the two important observations:

➢ Are the 1st and 3rd principles independent?
    ▪ The answer is tricky: yes and no
    ▪ They are in their presentation

- But the approximations made in the math formulation of the GTR (cut to 2nd order) render the equivalence principle also approximate
- Higher-order differentials and gradients do differ
- This will become noticeable in extremely high-energy processes.

➢ Einstein's theory is not the final one (AE *dixit*)
- Mach's principle of general relativity is not fulfilled
- Einstein was the first to admit his theory was approximate and was convinced that someone would soon perfect it
- We are trying to do this now, a century later: S-T and f(R) theories, QG? etc.
- The symmetry-breaking paradigm could be useful

**4. Events of 1923 and Einstein's Six-Month Journey**

*4.1. Notable Events in the World, in 1923*

Developing further the proposal sketched in Section 1, we shall now put into context Einstein visit to our country. To start, we summarize the most notable ephemerids of the year 1923. On 9 January, Juan de la Cierva made the first flight in his autogiro (a precursor of the helicopter). On the 11th, despite strong protests from the British, troops from France and Belgium occupied the Ruhr region to force Germany to pay them the war reparations that had been agreed upon in the Versailles Treaty. Already entering the month of February, we find in the international press the worrying news that, in Germany, inflation seems to have no ceiling: one dollar is exchanged for 57,500-mark notes. On 23 February, just as Einstein set foot in Barcelona, the German Parliament approved a decree law against speculators. However, hyperinflation in the Weimar Republic (Germany, at the time) continued to rise. In July, the number of mark notes needed to buy one American dollar reached 353,000, more than 200 times the amount needed at the start of 1923. And, on 15 November, hyperinflation became dramatically present in Germany and reached its peak: one dollar was exchanged for 4,200,000,000,000-mark notes (4.2 trillion!). These unbelievable figures have gone down in history as a record. This information must be kept in mind when we talk about Einstein's trip and visit to Spain and other circumstances that will be exposed next.

On 13 February 1923, Tutankhamun's tomb was discovered in Egypt. On 10 March of that year, the anarchist Salvador Seguí was murdered in Barcelona and, on 20 July, in Mexico, the popular leader Pancho Villa. On 13 September, the *coup d'état* led by General Primo de Rivera took place in Spain, which suspended the Constitution, dissolved the Parliament, and established the country's first dictatorship of the 20th century. On 16 October, Walt Disney and his brother Roy, with animator Ub Iwerks, founded Disney Bros. And, already towards the end of 1923, on 9 November, in Germany, the attempted *coup d'état* in Munich, known as the Beer Hall Putsch, failed. For this, Adolf Hitler and Rudolf Hess were shortly after prosecuted and sentenced to prison. We know well what happened later. This was the atmosphere, the air that was breathed in that year of 1923.

*4.2. Einstein's Long Journey*

Focusing now again on Einstein's activity, several very valuable publications record in all detail the long journey he undertook, accompanied by his wife at that time, Elsa Einstein, from the beginning of October 1922 to the end of March 1923, and which brought him to lecture at the Far East, Palestine, and Spain [32]. Those were places that the, already by then, very renowned physicist had never visited before. Einstein's long itinerary included stops in Hong Kong and Singapore, two short stays in China, a six-week lecture tour of Japan, a twelve-day tour of Palestine, and a three-week visit to Spain. Much more than a simple curiosity to see the world and make himself known personally, by giving talks everywhere, Einstein's trip responded to the purpose of getting away for a prudent time from Berlin, where the German nationalists had murdered, shortly before, the philosopher and Jewish diplomat Walther Rathenau. The brutality of his death, which happened while he was sitting in his car, in the street, due to the explosion of a hand grenade, had greatly impressed Einstein. He knew, moreover, that he and his wife were on "a list" and that it suited them to leave the country no matter what. We have the complete diary that Einstein wrote during those days [33]. To see the character of his narrative, here is a short extract from its first page:

**Travel diary for Japan, Palestine, and Spain [6 October 1922–12 March 1923].**

6 October *Night trip in overfilled train after reunion with Besso and Chavan. Lost wife at the border.*

7 October *Sunrise shortly before arrival in Marseille. Silhouettes of austere flat houses surrounded by pines. Marseille, narrow alleyways. Voluptuous women. Vegetative living. We were taken in tow by seemingly honest youth and dropped off at a ghastly inn by the railway station. Bugs in morning coffee. Made our way to the shipping company and the old harbor near the old city quarter. At the ship . . .*

According to Walter Isaacson's well-documented biography, entitled *"Einstein: his life and Universe"* [34], Mr. Koshin Morobushe Kaizosha, who was Einstein's Japanese host and publisher at the time, offered him the equivalent of two thousand pounds sterling (which would be about 150,000, at current exchange) for a series of lectures. They finally turned out to be fifteen, eight of which were scientific and six public, plus a memorable talk with students at Kyoto University, not previously planned.

It was on board the ship, during the trip that took him and his wife to Asia, when Einstein, then 43 years old, learned that he had been awarded the Nobel Prize in physics. He could not accept the prize in person at the Nobel ceremony in Stockholm in December 1922. On his behalf, the banquet speech was delivered by the German ambassador, who praised Einstein not only as a scientist but also as a man of peace and international activist. It was after the return of his long journey when, finally, on 11 July 1923, in Gothenburg, Einstein was able to deliver his speech in person. It was on the occasion of the meeting of the Scandinavian Scientists in an impressive auditorium and with the King of Sweden, Gustav V, sitting in the first row (Figure 8). Einstein chose to speak on the theory of relativity, even though the prize had not been formally awarded to him for this subject.

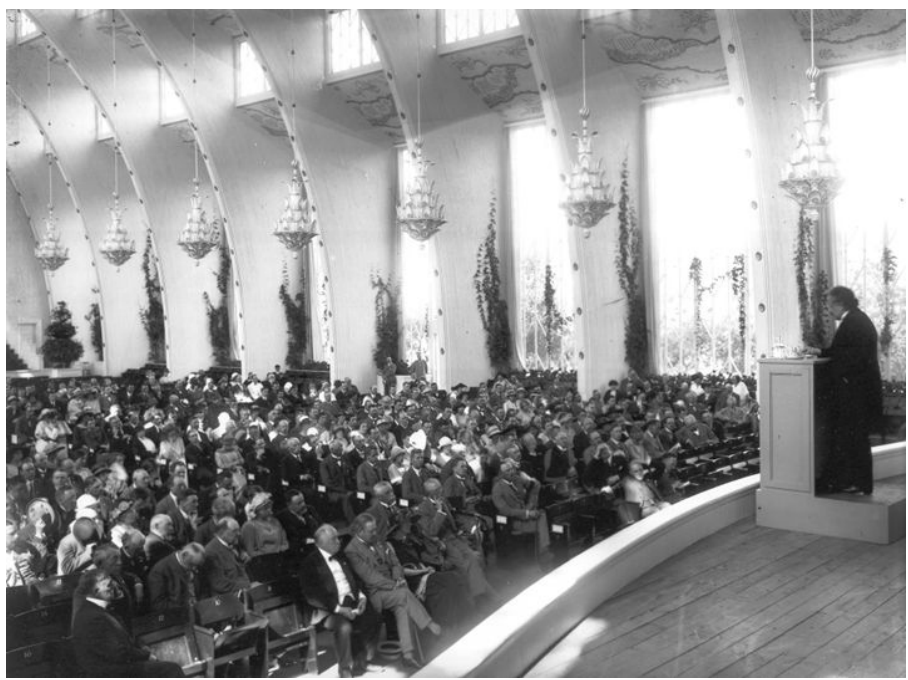

**Figure 8.** On 11 July 1923, Einstein spoke at the Congress Hall in Gothenburg, Sweden, at the Scandinavian Congress of Naturalists. Public domain.

He and his wife Elsa visited Japan from 17 November to 29 December 1922. His trip, meticulously organized by Kaizosha Publishing, made big international news. Japan was, in fact, the most important stop on his full tour, a tour that involved considerable effort for him. His thoughts, reflected in the meticulous personal diary Einstein kept, reveal a man trying to understand cultures very different from his own (Figure 9). His observations begin as early as they boarded the ship S.S. Kitano Maru, manned by a predominantly Japanese crew. It is clear from his notes that Einstein did not have much sensitivity when describing people from other cultures. On the ship, he sees, in his description, *"Japanese women crawling on deck with children"*; he sees them *"adorned and bewildered, almost as if they were sketchy, stylized, black-eyed, black-haired, big-headed, running..."*. Just before docking in Kobe, the ship stopped in Shanghai, and Einstein describes in his diary his frustration with Asian cuisine: *"The food, extremely sophisticated, endless. It is fished constantly, with sticks, of common bowls placed on the table in great numbers"*.

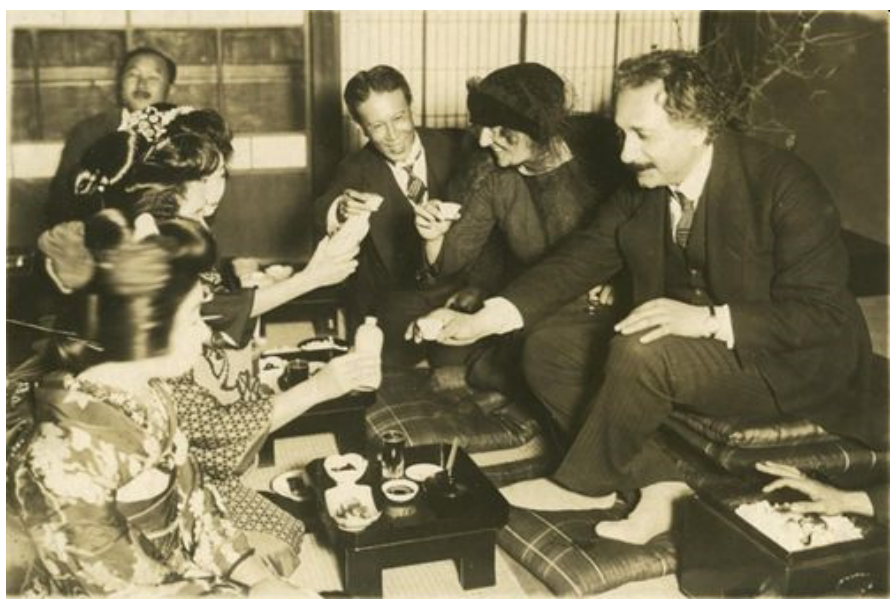

**Figure 9.** Albert and Elsa Einstein in Japan, November–December 1922. Author unknown, courtesy of Meiji Seihanjo. Public domain.

Upon arriving in Japan, Einstein was given a literal hero's welcome, and at times the fame and excessive attention overwhelmed him. One day on his tour, he was looking out the window just before sunrise. Below, thousands and thousands of Japanese were gathered, in vigil in front of the hotel. He shook his head and commented to Elsa: *"No living person deserves a reception like this. I see us as scammers. We will end up in prison"*. Both spent Christmas in Fukuoka, but mostly they toured the island of Honshu, making stops in Kobe, Kyoto, Tokyo, Sendai, Nikko, Nagoya, Osaka, Nara and Hiroshima. In Tokyo, Einstein felt, once again, suffocated by the attention: *"Arriving at the hotel, completely exhausted, among gigantic crowns of bouquets. Still to come: visit by the Berliners and live burial"*. To complete the day, he had been invited to the burial ceremony of a Japanese personality.

His scientific lectures were held at the Todai Institute of Physics and Tokyo Imperial University. Lasting four hours, the content was challenging for many aspiring Japanese scientists. As it had been before and would always be the case everywhere he imparted them; also, in particular, in the case of the several lectures he gave later in Barcelona, Madrid and Zaragoza, as should be obvious. In a letter dated 17 December 1922, he confessed to his sons:

> *"The Japanese people attract me... even more than all the peoples I have met so far: quiet, modest, intelligent, appreciative of art and considerate. Nothing is about appearances, but everything is about substance..."*

Since his comments were purely personal and not meant to be published, the reader now gets a clear look at Einstein's thinking process. The diary is a revelation of Einstein's mind at that time because it reflects his way of thinking outside of physics. However, it also contains several revealing notes on this matter, which we will discuss later.

Another example: on Christmas Day, in Fukuoka, Einstein traveled to the Moji YMCA (Figure 10), where he was photographed *"ten thousand"* times and felt lifeless: *"I was dead and my corpse back to Moji where it was dragged to a children's Christmas and had to play the violin"*. As he describes in his diary, exhausted, Einstein played *"The Ave Maria before collapsing at ten o'clock at night"*.

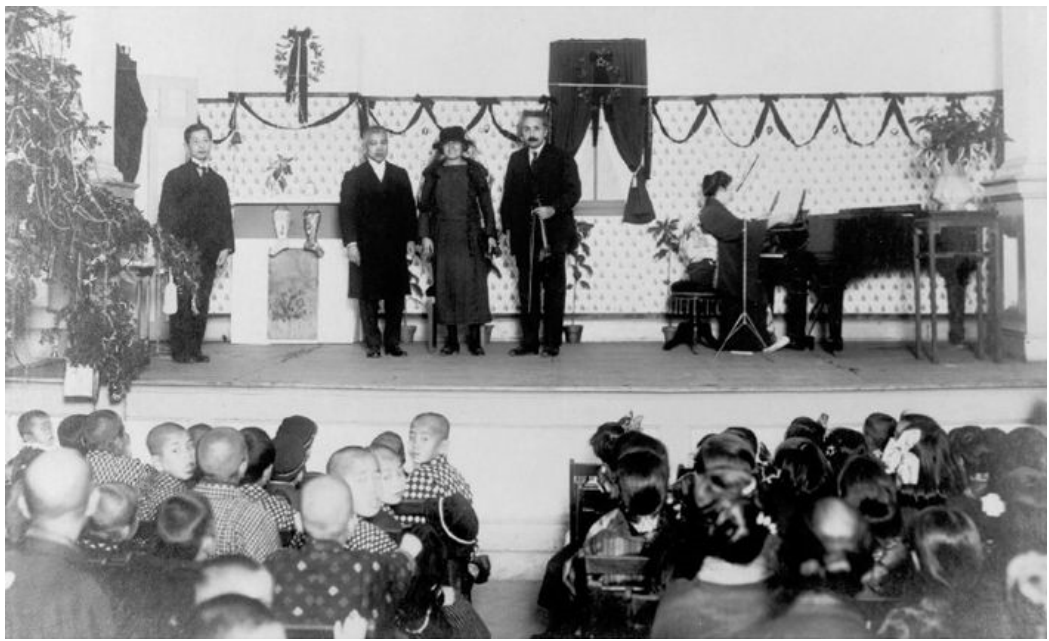

**Figure 10.** Einstein at the YMCA in Moji, Japan, December 1922. Courtesy of Kenji Sugimoto. Einstein Archives Collection, Hebrew University of Jerusalem. Fair use.

His last stop in Japan was the great city of the south, Hiroshima, which deserves special consideration. He arrived there by train on 19 December, close to the end of his stay. The previous day, the Einstein had been sightseeing in Nara, the country's old capital, from where they had taken the train for a twelve-hour journey. The next day, after recovering, Einstein took a *"fascinating walk along the coast"* of Miyajima and saw Itsukushima Shrine, one of Japan's three holiest sites. In the afternoon, he walked *"to the top of the mountain that gives the island its main shape"*, Mount Misen. It takes a few hours to reach the top, where he wrote, he saw *"the subtlest of colors"*, as well as

> *"...countless small temples, dedicated to natural deities. Stone figures are often delightful. Of steps cut in granite rocks (altitude around 700 m). Memorial to the Japanese love of nature and all kinds of endearing superstitions."*

On the beautiful summit of Mount Misen, Einstein was surrounded by *"pure souls like nowhere else in the world. One cannot but love and admire this country"*.

One may imagine that, from there, he could have glimpsed the center of Hiroshima and the Prefectural Industrial Promotion Hall, now known as the Peace Memorial or Genbaku dome (for the atomic bomb). Twenty-two years later, on 6 August 1945, when the atomic bomb detonated over the city center of Hiroshima, the explosion wave must have shattered the windows of the houses on the island that Einstein was admiring. His famous formula, $E = mc^2$, had led Lise Meitner, Otto Hahn, and other scientists [35], as discussed before, to investigate whether it could make real sense. And later, this led to advances in creating a chain reaction of uranium and plutonium and its use in the Second World War.

Although this had never been his intention, the moment Einstein heard the tragic news, he exclaimed: *"Woe is me!"*.

After a very demanding journey through Japan (Having personally been, thirty-five years ago, to the south of Japan (specifically, to the university of Hiroshima), in the capacity of guest as a visiting scientist, it is not difficult for me to take charge of the feelings of Einstein and the overwhelming reception they gave him. In my case, I was greeted with a spectacular firework display that ended with a thunderous crash, followed by a banquet with very select and plentiful food and drinks. The students, above all, were very happy about my visit, which gave occasion to such a great celebration, since these took place very sporadically at that time. I have been back there several times, but now everything has changed a lot. During my stay in Japan, I followed an itinerary almost identical to Einstein's and I feel very identified with what he says, with the deep impression he was made by Miyajima, for example. Those are enchanted sites and places that touch your heart.), [36] on 29 December 1922, Einstein and his wife set sail for the British protectorate of Palestine, where they arrived a month later (Figure 11). He had already started planning the visit to the Jewish community in those territories in 1921, but it had not been confirmed until shortly before he left Berlin for Japan. He would therefore combine the two stays, which became three, with the subsequent visit to Spain, also planned in 1921 and scheduled as the last stage of the long journey.

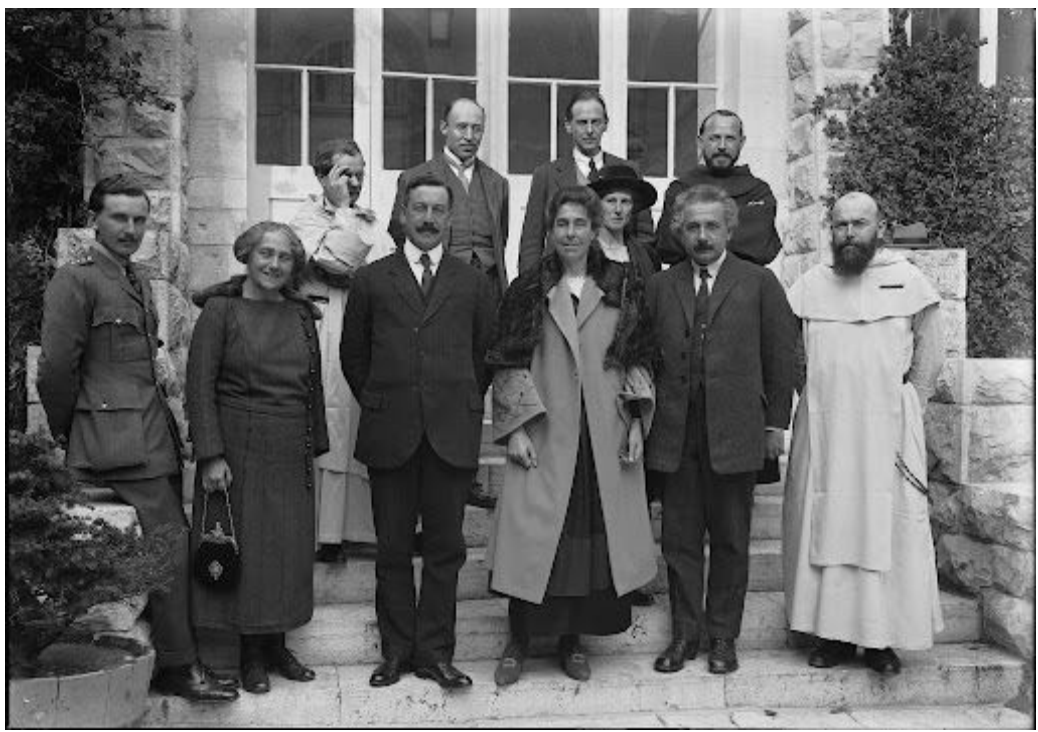

**Figure 11.** The Einsteins at the Government House in Jerusalem, with the British High Commissioner. February 1923. Einstein Archives Collection, Hebrew University of Jerusalem. Fair use.

In Palestine, Einstein stayed for twelve days, visiting the entire territory, the most important cities, and various agricultural settlements, as well as the main economic, social, and cultural institutions of the Jewish community. He also interviewed Arab leaders and the Christian community. In this case, Einstein's diary offers a priceless insight into the Jewish community in Palestine then. During those twelve days, Einstein visited Jerusalem, Tel-Aviv, and Haifa, traveled to the Dead Sea and Galilee, and lectured on relativity at the site that would later become the home of the Hebrew University.

*4.3. Einstein in Spain*

Einstein had been invited to visit our country by Esteve Terradas, a physicist like himself, working in Barcelona, and Julio Rey Pastor, a mathematician, working in Madrid. Independently, he had got an invitation to visit Zaragoza, as well. Terradas had offered him 7000 pesetas—in those days, twice the annual salary of a university professor—to give lectures in Barcelona and Madrid (Figure 12).

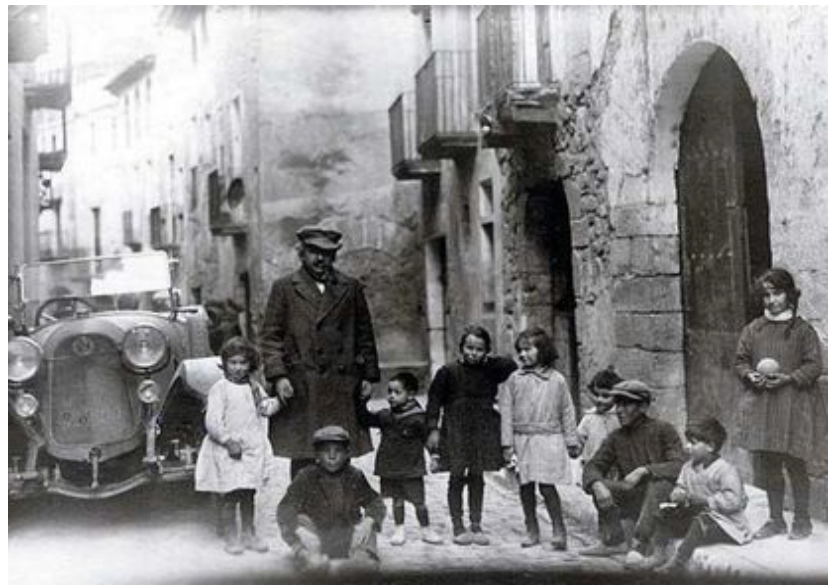

**Figure 12.** Albert Einstein in front of the Fonda Ibérica, in Espluga de Francolí, on 25 February 1923. What attracted the children's interest was not Einstein's charm but the magnificent automobile in which he had arrived, from Casa Elizalde, a type 29 torpedo, unmistakable (in 1922, its price was 33,000 pesetas)—public domain.

Returning from their stay in Palestine, the ship had docked at Marseilles where, as he carefully explains in his diary, Einstein had trouble checking the main part of their luggage directly to Berlin or Zurich failing that. After resolving the mishap, they set off by train for Barcelona, where they arrived on 22 February 1923. They had been unable to tell their hosts which train they had taken, and no one came to meet them at the station. With his wife Elsa, he headed to the modest hotel Cuatro Naciones, which is at the end of the Ramblas, to the left, going down. Numerous anecdotes refer to Einstein's humility, and this one is among the most famous. But they did not spend a single night there: when the owner of the hotel found out who Einstein was, he immediately told him that such was no place for him and his wife and sent them to the higher-class Colon hotel, located then in Catalunya Sq., corner with Passeig de Gràcia, where they spent the seven days of their visit. The stay at the Colon cost 692 pesetas. It was paid for, together with other expenses (including food and a bouquet), mounting to 883 pesetas by the city council of Barcelona. The entrance to each of his conferences cost 25 pesetas, quite a high price at the time; despite that, the classrooms were always full. We leave more details like those to expert historians with whom we do not intend to compete. It is interesting to read first-hand Einstein's notes corresponding to the days of his visit [37].

**Doc. 379. Travel diary [March 1923], p. 325–326**

"17, 18, 19 February. *Indigestion from bad food. High seas and rain. 19 in the morning, Stromboli well in sight. Afternoon, 6 o'clock, Naples. Vesuvius with gray clouds cloudy sky. So cold and unpleasant that one is glad to stay on the boat. An Englishman from Australia turns out to be from Mecklenburg. News of a rail strike in France and more and more retaliation in the Ruhr, how will things go? In Toulon, friendly people. In Marseille,*

> *dangerous to speak German. The manager of the freight depot refuses to send our baggage to Berlin or even to Zurich.*
>
> 22–28 February. *Stop in Barcelona. We are tired but friendly people (Terradas, Campalans, Lana, Tirpitz's daughter). Popular songs, dances. Refectory. How beautiful it was!* (Figure 13)
>
> 1 March *Arrival in Madrid. Departure from Barcelona, farewell. Terradas, German consul with Tirpitz's daughter, etc.* (Figure 14)
>
> 3 March *First lecture at the university.*
>
> 4 March *Car ride with Kocherthaler—answer to Cabrera. I wrote the academy speech. Academy session in the afternoon chaired by the king. Magnificent speech by the president of the acad. Afterwards, tea in the society of artists. Ladies. You felt at home but in a very Catholic atmosphere.*
>
> 5th *In the morning. Honorary member of the Mathematical Society. Debate on general relativity. Lunch at Kuno's. Visit with Kuchal. A wonderful old thinker. Very sick, good conversation. Invitation to dinner in the afternoon from Mr. Vogel. He has a good heart, humorous pessimism.*
>
> 6th *Excursion to Toledo hidden through many lies. One of the best days of my life. Radiant skies Toledo is like a fairy tale. We were guided by an old enthusiast, who supposedly had written something important about Gra [El Greco]. Streets and market, city views, the Tagus with stone bridges, stone covered hills, lovely level cathedral, synagogue, sunset on the return trip with brilliant colors. Small gardens with views near the synagogue. Greco's magnificent fresco in a small church (burial of a nobleman) is one of the most profound images I have ever seen. Wonderful day.*
>
> 7th *Audience at noon with the King and Queen Mother. The latter shows that she knows about science. You realize that no one tells her what they think. The king, simple and dignified, I admire him for his ways. In the afternoon, the third university conference, a devoted audience that probably couldn't understand practically anything because the latest problems were being discussed. In the evening, a great reception at the home of the German envoy. The envoy and family are magnificent and modest people. Socializing is as heavy as ever.*
>
> 8th *Honorary doctorate. Spanish speeches with associated firecrackers. Long but with good content that of the envoy on German-Spanish relations but in genuine German. No rhetoric. Then, in the afternoon, visit with the technical students. Speeches and nothing more than speeches, but very meaningful. Talk in the evening. Then playing music at Kuno's. A professional (director of the conservatory), Poras [Bordás] played the violin exquisitely.*
>
> 9th *Excursion to the mountain and El Escorial. Glorious day. An evening reception in the student residence with talks by Ortega and me.*
>
> 10th *Prado (mainly looking at paintings by Velazquez and El Greco). Farewell visits. Lunch at the home of the German envoy. A night with Lina and the Ullmanns in a small and primitive dance venue. Fun evening.*
>
> 11th *Prado (splendid masterpieces by Goya, Raphael, Fra Angelico).*
>
> 12th *Trip to Zaragoza."*

Finally, Einstein also visited Zaragoza, where he stayed from the 12th to the 14th of March and gave two lectures. The day he took the train back to Berlin, he turned 44.

Regarding Catalonia, it should be mentioned that he visited (in addition, obviously, to Barcelona) Sant Cugat del Vallès [38], Terrassa, Espluga de Francolí and Poblet. In his diary, the contrast between the only three or four lines dedicated to the stay in Catalonia and the more than forty that occupy the notes corresponding to the rest of the trip is evident. It must be said, however, that after the notes on Catalonia, he had left a blank page. Everything suggests that he intended to fill it later, which, unfortunately, he never did.

After the three-week visit, Einstein and his wife returned by train to Berlin, where they arrived on 21 March 1923, thus putting an end to their long journey. They had been out of Germany for nearly six months.

Historians, such as Thomas Glick, author of the highly recommendable reference book [39] *"The Spaniards and Einstein"* or also Ana Romero de Pablos, co-author of the book [40] *"Einstein en España"*, point out in a very similar way, that Einstein's visit did not serve to Europeanize Spanish science, nor did it open up new lines of research; they agree that people were left with, at the very least, a great feeling of admiration for the genius. Regarding its impact in Barcelona, in particular, it is good to read the article by Antoni Roca Rossell, *"Albert Einstein in Barcelona,"* in the work *"History, politics, society and Culture of the Catalan Countries"* and other publications (all in Catalan). More information about the visit to Catalonia and much of what has been written here can be found on the websites [41–47]. There is also an interesting "Einstein trail" through Barcelona [48].

One of the main objectives of the commemoration in memory of the centenary of Einstein's visit was to expose how this situation has changed radically. In a few words, in the Century that elapsed, Spanish scientists have transitioned from being passive admirers of scientific advances [39] to becoming actors at an international level who actively participate in cutting-edge scientific projects. Reliable evidence has been presented at conferences and scheduled events and can be obtained from international databases.

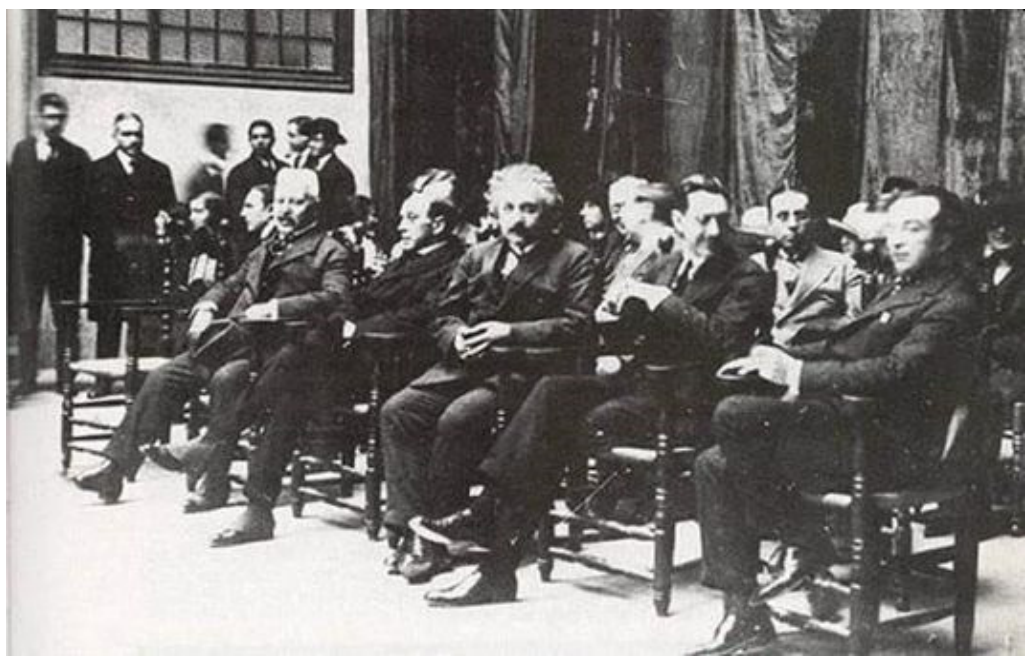

**Figure 13.** Einstein at the event held at the Industrial School of Barcelona on 28 February 1923, where he attended the performances of the Barcelona sardanist couple and the Penya de la Dansa of the New University Student Association. Public domain.

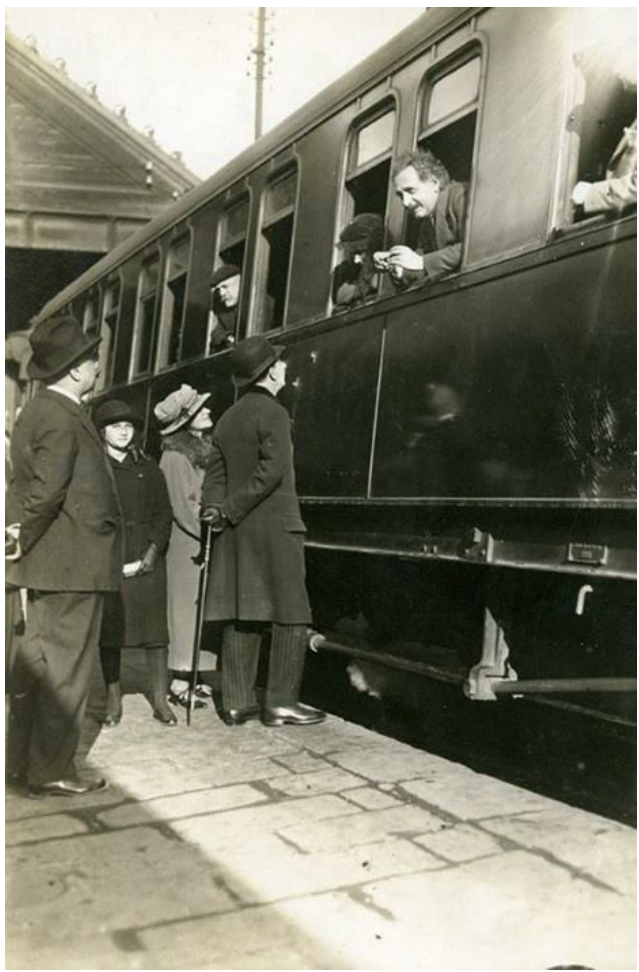

**Figure 14.** Einstein inside the train, at the France station in Barcelona, before leaving for Madrid on 1 March 1923. Public domain

## 5. The Important Scientific Context Around Einstein at the Time of His Visit

As it becomes clear from reading Einstein's travel diary, his mind was regularly occupied with questions having to do with physics. For example, on 9 October 1922, while on the ship that was taking him to Japan, he wrote in his diary that he was reading Ernst Kretschmer's book "Physics and Character" [49], as well as one of Henri Bergson's on relativity [50]. He explains his reflections on the matter in detail, filling more than one page with brief annotations. In particular, he says he is comparing the approaches of Riemann and Weyl to the problem of the unification of gravity with electricity, among other issues. He spoke about this subject at one of the conferences he gave, both in Barcelona and Madrid [39]. And it is quite well known that, already at that time and for several years of the rest of his life, Einstein devoted much of his time to trying to find a theory to unify gravitation and electromagnetism.

But it is not, in any case, this issue (although very important and still to be resolved) that will be described here, but rather, scientific events in a very different area, which took place in those days, around Einstein himself and his general theory of relativity (although he was not, this time, the hero of the confrontation). Those discoveries revolutionized our understanding of the Universe radically and eventually resulted in the creation of modern cosmology.

Alexander Friedmann (also sometimes spelt Alexandr Fridman) was born in 1888 in Saint Petersburg, where he remained for much of his short life. His father was a composer, and his mother was a dancer and pianist. He earned his bachelor's degree from St. Petersburg State University in 1910 and later became a professor at the city's Mining

Institute. Friedmann had acquired a great interest in the mathematics used in Einstein's general theory of relativity. Although already published a few years earlier, this theory was still not widely known in Russia due to the situations experienced in the First World War and, subsequently, the bloody Patriotic Revolution.

Friedmann was a friend of Paul Ehrenfest. They had known each other during the five-year stay of the last in St. Petersburg. Towards the end of 1920, Friedmann wrote Ehrenfest a letter in which he said:

> "...I have been working on the axiomatics of the principle of relativity, starting from two propositions: a) uniform movement continues to be uniform for all observers; b) the speed of light is constant (the same for both a static and a moving observer). Moreover, I have obtained formulas for a Universe with only one spatial dimension, which are more general than the Lorentz transformations..."

The Ehrenfest archive at the Lorentz Institute in Leiden (The Netherlands) also contains other letters and manuscripts that Friedmann sent to Ehrenfest, starting in early 1922. The translation of a letter he wrote to him, in Russian, in April of that year says:

> "...I am sending you a short note on the shape of a possible Universe, more general than Einstein's cylinders and De Sitter's spheres. Apart from these two cases, a world also arises whose space has a radius of curvature that varies with time. I thought this question might be of interest to you. As soon as I can, I will send you a German translation of this note. And, if you think the matter is interesting, please be so kind as to endorse me with a view to its publication in a scientific journal..."

This paper, "К Вопросу о геометрии Кривых пространств" ("On the Question of the Geometry of Curved Spaces"), dated 15 April 1922, does not appear in the surviving list of Friedmann's publications, which suggests that it was never published. Ehrenfest is known to have sent the manuscript—along with an (undated) letter Friedmann had written to Hermann Weyl—to the mathematician Jan Schouten, who was working in Delft. Schouten replied to Ehrenfest in a letter dated 29 June 1922, in which he criticized Friedmann's analysis (which did not prevent Friedmann and Schouten from collaborating, a few years later, on another purely mathematical subject).

In the same year, 1922, and while all this was happening, Friedmann translated his article into German. He elaborated it further and changed the title to: "О Кривизне пространства" ("On the Curvature of Space"). Now he introduced more clearly the idea of a possible curvature and expansion of space and decided to send it directly to the important journal Zeitschrift für Physik for publication.

The article was received by the journal on 29 June 1922. Friedmann demonstrated in it that the radius of curvature of the Universe could be an increasing or periodic function of time. He commented on the results of this paper in a book he wrote later, explaining them as follows:

> "...The case of a stationary universe includes only two possibilities, which have already been considered by Einstein and De Sitter. The case of a variable universe admits, on the other hand, many possible situations. In some cases, the radius of curvature of the universe increases steadily with time. And other situations correspond to a radius of curvature that changes periodically..."

Einstein analyzed Friedmann's paper quite quickly, as evidenced by the fact that Zeitschrift für Physik received his reply on 18 September 1922 [51] (a few weeks before embarking on the six-month-long journey described above):

> "...As for the non-stationary universe, the results contained in the work seem suspicious to me. The solution given for this case turns out not to satisfy the field equations..."

Friedmann learned of Einstein's criticism through his friend Yurii Krutkov, who was visiting Berlin then. And, on 6 December, Friedmann wrote a letter to Einstein responding to his objections:

> "...Considering that the possible existence of a non-stationary universe is of interest, I would like to present the calculations I have made here so that you can verify and critically evaluate them. [He details all mathematical operations]. If you find the calculations, I present in this letter to be correct, please be so kind as to inform the editors of Zeitschrift für Physik about this conclusion. Perhaps in that case, you would like to publish a correction to the statement you have made, or at least allow me to publish the calculations part of this letter..."

However, when Friedmann's letter reached Berlin, Einstein had already embarked on the long journey that took him and his wife to Japan, Palestine, and Spain. He did not return to Berlin, as we have seen before, until March of the following year. But, even after returning, Einstein did not read Friedmann's letter for the time being (or perhaps deliberately ignored it altogether, this is unknown).

Anyway, in May 1923, Krutkov and Einstein met again in Leiden, where both gathered there for the last master class of Hendrik Lorentz, who was retiring as a professor of his own will. They met, face to face, in the house of Ehrenfest, who was precisely the one who would succeed Lorentz in his chair. There, Krutkov could explain to Einstein the details in Friedmann's letter.

The result of the scientific discussion that subsequently took place is known through two short paragraphs of respective letters that Krutkov wrote to his sister in St. Petersburg a few days later. In the first, he explains:

> "...On Monday, May 7, I was with Einstein, reading Friedmann's article of Zeitschrift für Physik in detail..."

And finally, in the other letter, written on 18 May 1923, he states:

> "...I managed to defeat Einstein in the argument of Friedmann's work. Petrograd's honor is saved!"

Einstein had finally admitted his error and immediately wrote a note to Zeitschrift für Physik retracting his earlier observation:

> "...In my previous note, I criticized Friedmann's work on the curvature of space. However, a letter from Mr. Friedmann, which Mr. Krutkov handed me, has convinced me that my criticism was based on an error in my calculations. Now I consider that the results of Mr. Friedmann are correct and bring new light. It is shown that the field equations and the static solution also admit dynamic solutions (i.e., with a variable time-coordinate), with central symmetry for the spatial structure."

The retraction note [52] was received in Zeitschrift für Physik on 31 May 1923 (Figure 15). In any case, this did not mean, at all, that Einstein had become convinced that Friedmann's equations were of any use, that they could have anything to do with physical reality (although they appeared to be mathematically correct, at least).

Friedmann's equations, coming from Einstein's and corresponding to a homogeneous and isotropic universe, are given in terms of two expressions. The first one is derived from the 00 component of Einstein's field equations:

$$H^2 \equiv \left(\frac{\dot{a}}{a}\right)^2 = \frac{8\pi G \rho + \Lambda c^2}{3} - \frac{Kc^2}{a^2}.$$

The second one reads:

$$3\frac{\ddot{a}}{a} = \Lambda c^2 - 4\pi G \left(\rho + \frac{3p}{c^2}\right),$$

and comes from the first, together with the trace of Einstein's field equations (the dimension of the two equations being $T^{-2}$).

In these equations, a is the scale factor, G, Λ, and c are universal constants: namely, G is the Newtonian constant, Λ the cosmological constant with a dimension of $L^{-2}$, and c is the speed of light in vacuum. Moreover, ρ and p are the volumetric mass density and pressure, respectively; k, corresponding to the curvature, is constant throughout a particular solution but may vary from one solution to another.

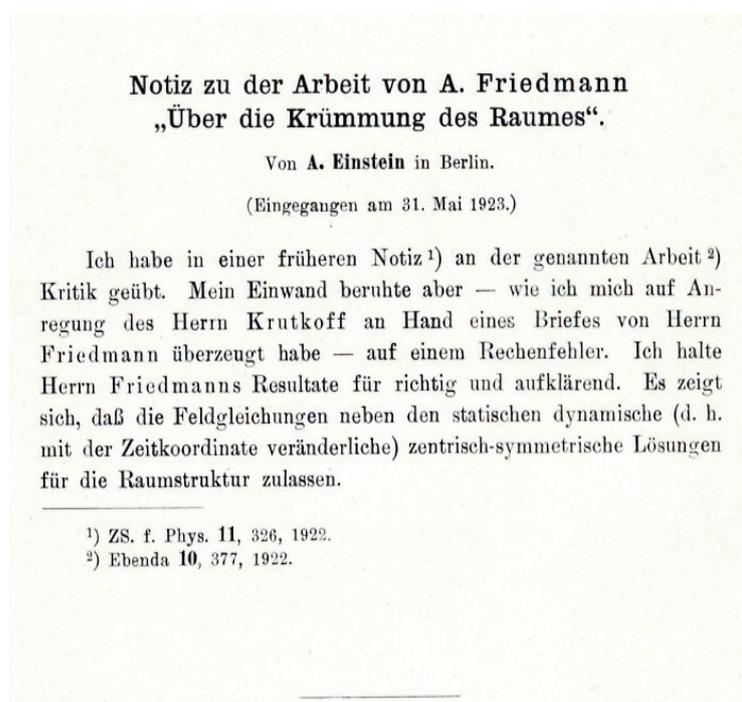

**Figure 15.** Retractation note by Albert Einstein, published in Zeitschrift für Physik on 31 May 1923—public domain.

It can be stated, however, that, for at least a decade, no one considered the works of Friedmann (he supplemented the first with a second one, published in 1924. see below) as possible models for our Universe. For instance, it took Einstein another ten years to admit the expansion of the Universe as an actual physical possibility, despite the astronomical evidence accumulating during that time [3,4].

On the other hand, there is an essential fact that is systematically overlooked. In his work, Friedmann obtained several families of equations, not only the one now known as *"Friedmann's"*, since he did not necessarily impose the restriction that the Universe should be homogeneous and isotropic. In addition to possible universes with an initial singularity and others that started with a finite radius, he found many more cases, including periodic solutions; and one with a logarithmic term, which implied that the Universe had no origin in time, extending from minus- to plus-infinity.

It was precisely a solution of this last type, the one that Lemaître discovered (or rediscovered) a couple of years later. On this, he based his first model of the Universe published in 1927 [53], thus getting ahead of all the geniuses of astronomy and theoretical physics by affirming resoundingly that the Universe *was* expanding, as demonstrated by the astronomical data given to him by Vesto Slipher and Edwin Hubble, which were in perfect agreement with the solutions to Einstein's equations that he had found—later than Friedmann, in any case. And we must observe again that in this first work, he did not use "the" Friedmann's equation, with an origin for time, the one accepted now.

To complete the information, in 1924, Friedmann published, as already mentioned, a second work [54], also in Zeitschrift für Physik: "*Uber die Möglichkeit einer Welt mit constanter* *negativer Krümmung des Raumes"* ('On the possibility of a world with constant negative space curvature'). This work completed the previous one from 1922, obtaining all possible cases for the values of the curvature of the Universe: positive, negative or null. Ten years later, Howard Robertson and Arthur Walker rigorously showed that if the Universe is homogeneous and isotropic (e.g., it satisfies the cosmological principle), the only family of Friedmann solutions that survives is the one that originates with a singularity—the curvature can still be positive, negative or zero.

To conclude, after spending years completely unnoticed, the dynamic cosmological model of general relativity, originating in Einstein's equations and completed by Friedmann, would become the only possible solution for our Universe. It can be affirmed that, in this way, the magnificent framework of general relativity finally became the theory *par excellence* for the description of our Universe. To the particular beauty in its conception from seemingly natural principles (which we will never tire of underlining) was now added the extremely important fact that it is the only possible theory (within the framework of the postulates, as established by Einstein) (Figure 16) and that it has a unique solution: the one found by Friedmann, now commonly called of Friedmann-Lemaître-Robertson-Walker (FLRW).

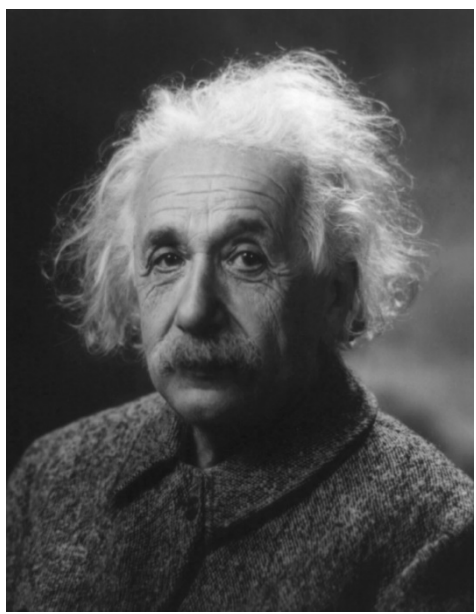

**Figure 16.** Albert Einstein in Princeton, in 1947. Public Domain.

A last note on Friedmann. In June 1925, he obtained the position of director of the Main Geophysical Observatory in Leningrad (the new name of his city), and, in July, he took part in a balloon flight that set an altitude record for the time of 7400 m. He died soon after, on 16 September 1925, at the age of 37, of a misdiagnosed typhoid fever; which he contracted on his way back from his honeymoon in Crimea after eating an unwashed pear he had bought at a railway station.

## 6. Summary and Conclusions

What we have described in this article is a crucial development in the middle of what is now known as the first revolution in modern cosmology. A revolution that the author has framed in the twenty years from 1912 to 1932. It started with the transcendental astronomic discoveries made by Henrietta Leavitt and Vesto Slipher and extended up to those of Edwin Hubble; and included the no-less important theoretical advances of Albert Einstein, Alexander Friedmann, Willem de Sitter, and Georges Lemaître [3,4]. There is a consensus that it peaked in 1929 with the publication of Hubble's results. It is too often forgotten that it was Lemaître who, in 1927, had published Hubble's law, together with his perfectly reasoned and documented conclusions on the fact that the Universe was expanding. And this is what he told Einstein in person that same year, at the famous Fifth Solvay Conference, held in Brussels. However, Einstein did not accept his conclusions and let him know that his physical intuition was "abominable". Eventually, the theory of the expansion of the Universe, having an origin in the past, was adopted by all the main specialists and was crowned with the famous model by Einstein and De Sitter in 1932.

Notwithstanding that, the final confirmation of the Universe's expansion as a true scientific theory had still to wait for a very elaborate formulation of the Big Bang model, its definitive verification through the detection of the cosmic background radiation (CMB), and yet a major and crucial reshaping (inflation), which would only arrive fifty years later. And which was already the prelude to a second revolution (1985–2005) [3]: The expansion of the Universe is accelerating. To wit, according to the most recent and accurate astronomical observations, it is very likely that our universe had an origin from (almost) nothing (e.g., from a vacuum state of a tiny quantum system including space-time and a scalar field or two) some 13.8 billion years ago and is currently in accelerated expansion. No one has been able to explain this last fact convincingly yet, which has turned into one of the main open problems of present-day cosmology. All attempts to keep Einstein's general relativity and explain the acceleration using a (possibly running) cosmological constant—that would come from feasible contributions of quantum vacuum fluctuations—have failed to date.

Anyway, Einstein's theory will have to be modified, this is for sure. We noted already that it was Einstein himself the first to recognize this fact, having failed to incorporate Mach's ideas properly. He dared to preview that some of his colleagues would improve his equations soon. More than a century later, theoreticians are still working on this issue, mainly because of the need to cope with experiments and observations on two different playgrounds, which are, in fact, closely connected. On one side, on small distances, e.g., the realm of quantum physics and beyond, up to the GUT's scale or the inflationary one. And, on the other, in the realm of black-hole collisions and other extremely energetic processes in the universe, such as GRBs and others. In the absence of a quantum theory of gravity, perturbation terms under the form of powers and other functions of the curvature are being added to the original Einsteinian, second-order theory, using convincing arguments of different sorts (What we can also understand as going in the direction of trying to fulfill Mach's principle.). A lot of work is being done in those directions, and our research group in Barcelona and different collaborators have issued pioneering papers on some of these subjects (see, e.g., Refs. [55–64]).

What we have just described above constitutes, without a doubt, a pivotal episode in the history of physics, cosmology and, even further, in all human history. And, as we can appreciate—this was the purpose of the last section—a crucial act of this episode occurred in 1922–1923, around the time of Einstein's visit to our country.

Although we cannot say that his contribution to this issue was as brilliant as in many other cases (Section 2), we should appreciate that the fundamental basis for the whole discussion continues to be the field equations of his general theory of relativity, conveniently investigated further by other researchers of great insight and intuition.

Finally, it is a fact that Einstein himself could not grasp all the consequences of the exceptional theory he had created, starting from a few very basic and natural principles. It has taken more than a century and the dedication of thousands of researchers worldwide to get an extended idea of them. This shows us, palpably, that, despite the importance of the great geniuses at times may appear to be infinite, progress in knowledge is always, without exception, a collective task.

**Funding:** This research was funded by the Spanish State Research Agency program AEI/10.13039/ 501100011033, project number PID2019-104397GB-I00, by AGAUR, Catalan Government, project 2017-SGR-247, and by the program Unidad de Excelencia María de Maeztu CEX2020-001058-M.

**Data Availability Statement:** Not applicable.

**Acknowledgments:** This paper is based on the author's opening talk at the Fourth International Conference on Symmetry and two more talks at the Royal Academy of Sciences and Arts of Barcelona and the Institute of Space Sciences in Bellaterra. Comments from the participants in these events and very helpful observations from two manuscript referees are gratefully acknowledged.

**Conflicts of Interest:** The author declares no conflict of interest.